\documentclass[aip,cha,graphicx,amsmath,amssymb,reprint]{revtex4-2}

\usepackage{graphicx}
\usepackage{dcolumn}
\usepackage{bm}
\usepackage{acronym}
\usepackage[utf8]{inputenc}
\usepackage[T1]{fontenc}
\usepackage{mathptmx}
\usepackage{etoolbox}
\usepackage{amsmath}

\usepackage{hyperref}

\begin{document}

\title{Rydberg excitons in cuprous oxide: A two-particle system with classical chaos}
\author{Jan Ertl}
\author{Sebastian Rentschler}
\author{Jörg Main}
\email[Email: ]{main@itp1.uni-stuttgart.de}
\affiliation{Institut für Theoretische Physik I,
  Universität Stuttgart, 70550 Stuttgart, Germany}

\date{\today}

\begin{abstract}
  When an electron in a semiconductor gets excited to the conduction band
  the missing electron can be viewed as a positively charged particle,
  the hole. Due to the Coulomb interaction electrons and holes can form
  a hydrogen-like bound state called exciton.  For cuprous oxide a Rydberg
  series up to high principle quantum numbers has been observed by
  Kazimierczuk et al.\ [Nature \textbf{514}, 343 (2014)] with the
  extension of excitons up to the µm-range.  In this region the
  correspondence principle should hold and quantum mechanics turn into
  classical dynamics.  Due to the complex valence band structure of
  Cu$_2$O the classical dynamics deviates from a purely hydrogen-like
  behavior.  The uppermost valence band in cuprous oxide splits into
  various bands resulting in a yellow and green exciton series.  Since
  the system exhibits no spherical symmetry, the angular momentum is
  not conserved.  Thus, the classical dynamics becomes non-integrable,
  resulting in the possibility of chaotic motion.  Here we investigate
  the classical dynamics of the yellow and green exciton  series in
  cuprous oxide for two-dimensional orbits in the symmetry planes as
  well as fully three-dimensional orbits.  The analysis reveals
  substantial differences between the dynamics of the yellow and green
  exciton series. While it is mostly regular for the yellow series
  large regions in phase space with classical chaos do exist for the
  green exciton series.
\end{abstract}


\maketitle

\acrodef{PSOS}[PSOS]{Poincar\'e surface of section}
\acrodefplural{PSOS}[PSOS]{Poincar\'e surfaces of section}
\acrodef{LD}{Lagrangian descriptor}

\begin{quotation}
In this paper we investigate the classical dynamics of the yellow and green exciton
series in cuprous oxide. Due to the crystal environment the symmetry 
is drastically reduced compared to the hydrogen-like case and additional 
terms consistent with the cubic O$_\mathrm{h}$ symmetry appear when describing the 
band structure. Through an adiabatic approach we arrive at a description
of the classical dynamics via energy surfaces in momentum space, which 
can be assigned to the different series. Classical orbits can be
calculated by choosing the corresponding energy surface and integrating 
Hamilton's equations of motion. 
Due to the cubic symmetry fully three-dimensional motion is possible
in addition to the two-dimensional motion in the two distinct symmetry
planes of the crystal with $C_{4v}$ and $C_{2v}$ symmetry.
We investigate the stability
properties of orbits and analyze the classical exciton
dynamics by application of Poincar\'{e} surfaces of section and
Lagrangian descriptors.
The analysis reveals the existence of a classically chaotic exciton
dynamics for both yellow and green excitons, however, the chaotic
regions are more pronounced for the green than for the yellow excitons.
\end{quotation}

\section{Introduction}\label{sec:introduction}
In contrast to the weak and strong interaction, which only become important
at subatomic length scales, the fundamental forces of 
gravity and electrodynamics mediate long-range interactions, where the
gravitational as well as the Coulomb force have the same $1/r^2$ dependence.
As such, they play the dominant role from atomic to astronomic length scales.

The simplest is the interaction of two particles.
In astronomy the gravitational force between two bodies leads, in case
of bound motion, to orbits on Kepler ellipses.
Due to the similar structure the same holds true in atomic physics, 
where Bohr introduced a quantization scheme which was later 
expanded by Sommerfeld, allowing to recover the quantum spectrum by assigning
Kepler ellipses to the corresponding quantum state.

In these two-particle systems the dynamics is completely regular, and
no chaotic motion is possible. This changes however when introducing a
third body, where chaotic dynamics become possible.
While historically Poincar\'{e}'s work on the three-body problem laid
down the foundation for chaos theory,\cite{barrow1997poincare} the
problem is still a topic of active
research.\cite{stone2019statistical,Breen2020Newton,liao2022three}

The dynamics of a two-particle system can change completely, when
adding external forces.
On atomic scales the hydrogen atom in a magnetic field is the prototype example 
of a quantum system with underlying classically chaotic dynamics and
thus for the study of quantum chaos.\cite{Friedrich1989The,Hasegawa1989Classical}

Excitons can be seen as the solid state analogue of the hydrogen atom.
In cuprous oxide they exhibit a Rydberg series up to high principle quantum numbers.\cite{Kazimierczuk2014Giant,versteegh2021giant}
However, a more detailed investigation shows a splitting of different $l$-states\cite{Schoene2016Deviations,Schoene2016Coupled}
and even a fine-structure splitting for different $l$-manifolds.\cite{Thewes2015Observation}
These deviations from the hydrogen-like behavior can be traced to the complex band 
structure of cuprous oxide, which has to account for the cubic
symmetry of the system.\cite{Schoene2016Deviations,Schweiner2016Impact}
This has been achieved in the Luttinger-Kohn theory\cite{Luttinger55,
  Luttinger56,Suzuki74}, where a quasispin $I=1$ is introduced.
The coupling between the quasispin
and the hole spin results in a splitting of the valence band and leads to
a yellow\cite{Kazimierczuk2014Giant,Schoene2016Coupled,versteegh2021giant} 
and green exciton series.\cite{Rommel2020Green,Schoene2016Coupled,versteegh2021giant,Grun1961Determination}

In a quantum mechanical framework the existence of quantum chaos has been demonstrated
for magnetoexcitons in cuprous oxide which break all antiunitary symmetries.\cite{Schweiner2017MagnetoexcitonsBreak,Schweiner2017goe,schweiner2017crossover}
The classical dynamics can be examined by following an adiabatic approach for the spin
degrees of freedom.\cite{Ertl2020Classical}
The resulting classical orbits can be connected to the quantum spectra by 
application of semiclassical methods.\cite{Ertl2022Signatures,Ertl2024}
Due to the cubic symmetry of the system angular momentum is no longer conserved.
In general this means that three-dimensional motion becomes possible and 
two-dimensional motion only is to be expected in the two distinct symmetry 
planes of the crystal.
The dynamics in the symmetry planes can be analyzed using a \ac{PSOS},
which fully characterizes the dynamics through one coordinate and the
corresponding momentum, while the remaining coordinates and momenta
are determined by an intersection condition and the conservation of
energy.\cite{lichtenberg2013regular}
Line shaped structures in the \ac{PSOS} indicate tori in phase
  space related to regular integrable or near-integrable classical
  dynamics, while regions in phase space with deterministic chaos are
  indicated by stochastic areas in the \ac{PSOS}.
For the yellow exciton series the majority of phase space exhibits regular behavior 
characterized by deformed tori in the \ac{PSOS}
and only small regions in phase space with deterministic chaos.\cite{Ertl2020Classical,Ertl2024} 
For three-dimensional motion the \ac{PSOS} is no longer suitable to depict 
the emerging phase space structure. However, other geometrical properties of the 
orbits can be used to characterize the dynamics.
An example is the \ac{LD}, which measures the arc length of an orbit 
for a given time interval and starting configuration.\cite{mendoza2010hidden,mancho2013lagrangian}
Recently it has been shown that an indicator based on the second
derivatives of the \ac{LD} provides a powerful tool for revealing the
global phase space structures of a dynamical system.\cite{Daquin2022Global}
In addition to these qualitative techniques one can also analyze the
Lyapunov stability of orbits.
The Lyapunov exponents provide a quantitative measure for the
sensitivity to initial 
conditions provided by the trajectories.

The paper is organized as follows.
After briefly discussing the theoretical description of excitons in cuprous oxide
in section~\ref{sec:theory}, we proceed by investigating the classical dynamics
of the yellow and green exciton series in section~\ref{sec:results}.
For two-dimensional orbits in the symmetry  planes of the crystal we
compare surfaces of section for the distinct series, revealing large
chaotic phase space regions for the green series.
To gain a better understanding of the fully three-dimensional motion
we use an \ac{LD}.\cite{mendoza2010hidden,mancho2013lagrangian,Daquin2022Global}
We investigate a selection of periodic orbits and their Lyapunov
exponents, for varying energy.
A change of the stability properties is observed, and related to
a pitchfork bifurcation breaking the initial symmetry of orbits.
We conclude with a summary and outlook in section~\ref{sec:conclusion}.

\section{Theory and methods}
\label{sec:theory}
\subsection{Excitons in cuprous oxide}
\begin{figure}
  \includegraphics[width=\columnwidth]{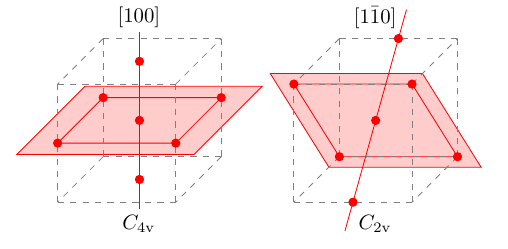}
  \caption{The two distinct symmetry planes of the $O_{\mathrm{h}}$
    group and the corresponding rotational axes normal to the plane.}
  \label{fig:symmetries}
\end{figure}

In a simple two-band approach one can consider electrons
in a semiconductor as nearly free particles
leading to a parabolic dispersion for the electron in the conduction band,
\begin{align}
  &H_{\mathrm{e}} (\boldsymbol{p}_{\mathrm{e}})
    = E_{\mathrm{g}}+\frac{1}{2m_\mathrm{e}} \boldsymbol{p}_{\mathrm{e}}^2
    \, ,
\label{eq:H_kin_e}
\end{align}
where $m_\mathrm{e}$ is the effective mass of the electron in the conduction band
and $E_{\mathrm{g}}$ the band gap energy.
A similar approach would lead to a parabolic dispersion  
also for the hole in the valence band
with the effective hole mass $m_\mathrm{h}$.
This two-band model assumes a homogeneous crystal background.
However, in reality the cuprous oxide crystal exhibits the 
cubic $O_{\mathrm{h}}$ symmetry,\cite{Koster1963}
which needs to be considered to accurately capture
the band structure of the system.
The $O_{\mathrm{h}}$ group contains all symmetry operations which map the cube back onto itself.
These include four threefold rotations around the $[111]$ axis and its equivalents,
six twofold rotations around the $[1\bar{1}0]$  axis and its equivalents,
as well as three fourfold rotations around the $[100]$ axis and its equivalents.
In addition, two distinct types of symmetry planes exist mapping the cube to 
itself via reflection.
There exist six mirror planes normal to the $[1\bar{1}0]$ axis which exhibit a
$C_{2v}$ symmetry and three mirror planes normal to the $[100]$ axis and its equivalents
with a $C_{4v}$ symmetry.
The symmetry planes of the crystal are visualized in figure~\ref{fig:symmetries}.

\begin{figure}
  \includegraphics[width=\columnwidth]{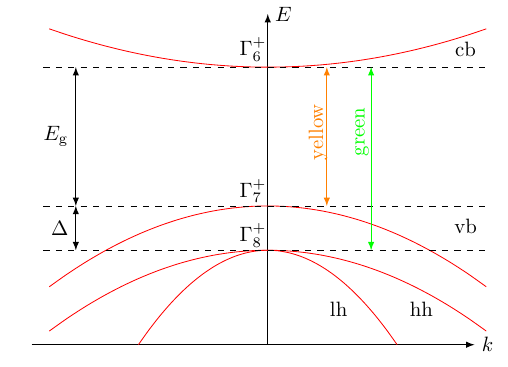}
  \caption{Schematic view of the band structure of Cu$_2$O. The
    valence band consists of a band with light holes (lh) and heavy
    holes (hh) as well as a split off band separated by the spin-orbit
    splitting $\Delta$. The light needed to excite an electron from
    the valence band (vb) to the conduction band (cb) belongs to the
    green and the yellow part of the spectrum, resulting in two green
    and a yellow exciton series.}
  \label{fig:bandstructure}
\end{figure}

While the parabolic dispersion captures the form of the lowest valence
band well (the conduction band stems from a copper $4s$-orbital and
belongs to the irreducible representation
$\Gamma_1^+$\cite{robertson1983electronic,Koster1963} leading to a
parabolic dispersion),
the simple two-band model is insufficient to accurately describe the
valence band structure of cuprous oxide.
The uppermost valence band in cuprous oxide stems from a copper $3d$-orbital, due 
to the cubic symmetry of the crystal it splits into a higher-lying
$\Gamma_5^+$-band and a lower-lying $\Gamma_3^+$-band.\cite{robertson1983electronic,Koster1963}
The $\Gamma_5^+$-valence-band can be theoretically modeled by introducing a quasispin $I=1$ 
to which the spin of the hole $S_\mathrm{h}$ in the valence band
couples via\cite{Luttinger55,Luttinger56,Suzuki74}
\begin{equation}
  H_{\mathrm{SO}}=\frac{2}{3}\Delta
  \left(1+\frac{1}{\hbar^2}\boldsymbol{\hat{I}}\cdot\boldsymbol{\hat{S}}_{\mathrm{h}}\right)\, .
\label{eq:H_SO}
\end{equation}
This leads to a splitting of the yellow and green series by the
spin-orbit coupling $\Delta$.
In addition to the parabolic dispersion all terms in accordance with the cubic symmetry 
need to be considered\cite{Schoene2016Coupled,Luttinger55,
  Luttinger56,Suzuki74}
\allowdisplaybreaks
\begin{align}
  &H_{\mathrm{band}} (\boldsymbol{p}_{\mathrm{h}},\boldsymbol{\hat{I}},\boldsymbol{\hat{S}}_{\mathrm{h}})
    = \frac{1}{2\hbar^2m_0} \big[4\gamma_2\hbar^2\boldsymbol{p}_{\mathrm{h}}^2\nonumber\\[1ex]
  &-6\gamma_2(p^2_{\mathrm{h}1}\hat{I}^2_1+{\rm c.p.})
    -12\gamma_3(\{p_{\mathrm{h}1},p_{\mathrm{h}2}\}\{\hat{I}_1,\hat{I}_2\}+{\rm c.p.})\nonumber\\[1ex]
  &-12\eta_2(p^2_{\mathrm{h}1}\hat{I}_1\hat{S}_{\mathrm{h}1}+{\rm c.p.})
    +2(\eta_1+2\eta_2)\boldsymbol{p}_{\mathrm{h}}^2(\boldsymbol{\hat{I}}\cdot\boldsymbol{\hat{S}}_{\mathrm{h}})\nonumber\\[1ex]
  &-12\eta_3(\{p_{\mathrm{h}1},p_{\mathrm{h}2}\}(\hat{I}_1\hat{S}_{\mathrm{h}2}
    +\hat{I}_2\hat{S}_{\mathrm{h}1})+{\rm c.p.})\big]+H_{\mathrm{SO}} \, ,
\label{eq:H_band}
\end{align}
with the Luttinger parameters $\gamma_i$ and $\eta_i$,
$m_0$ the free-electron mass, c.p.\ for cyclic permutation, and
$\{a,b\}=\frac{1}{2}(ab+ba)$ the symmetrized product.
The kinetic energy of the hole then consists of a parabolic part and
corrections due to the cubic symmetry of the system
\allowdisplaybreaks
\begin{align}
  &H_{\mathrm{h}} (\boldsymbol{p}_{\mathrm{h}},\boldsymbol{\hat{I}},\boldsymbol{\hat{S}}_{\mathrm{h}})
    = \frac{\gamma_1}{2m_0} \boldsymbol{p}_{\mathrm{h}}^2
    + H_{\mathrm{band}} (\boldsymbol{p}_{\mathrm{h}},\boldsymbol{\hat{I}},\boldsymbol{\hat{S}}_{\mathrm{h}})\, .
\label{eq:H_kin_h}
\end{align}
A schematic view of the band structure is shown in figure~\ref{fig:bandstructure}.
Excitons are formed when an electron in the conduction band and a hole
in one valence band are bound via the Coulomb interaction.
After transforming to relative and center of mass coordinates and neglecting the center of
mass motion the full Hamiltonian reads
\begin{equation}
  H_\mathrm{ex} = E_{\mathrm{g}} + \frac{\gamma'_1}{2m_0} \boldsymbol{p}^2
  -\frac{e^2}{4\pi\varepsilon_0\varepsilon|\boldsymbol{r}|} +H_{\mathrm{band}}(\boldsymbol{p},\boldsymbol{\hat{I}},\boldsymbol{\hat{S}}_{\mathrm{h}}) \, ,
\label{eq:H_spins}
\end{equation}
with $\gamma'_1= \gamma_1 + m_0/m_{\mathrm{e}}=2.77$\cite{Hodby1976Cyclotron,Schoene2016Deviations} and
$\varepsilon=7.5$\cite{Madeulung1998LandoltBornstein} the dielectric constant of Cu$_2$O.
The eigenvalues of the Hamiltonian~\eqref{eq:H_spins} can then be determined
using quantum mechanical
methods\cite{Thewes2015Observation,Schoene2016Deviations,Schoene2016Coupled,Schweiner2016Impact}
to describe the experimental spectra,\cite{Kazimierczuk2014Giant} however
without providing a physical interpretation of the states in terms of
classical dynamics.

\subsection{Adiabatic approach and classical exciton dynamics}
\begin{figure}
  \includegraphics[width=0.9\columnwidth]{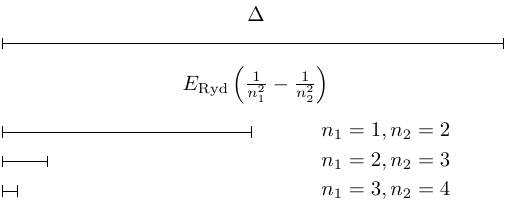}
  \caption{Comparison of the spin-orbit splitting $\Delta$ to the
    spacing between the first four Rydberg levels.}
  \label{fig:energy}
\end{figure}

\begin{figure}
  \includegraphics[width=\columnwidth]{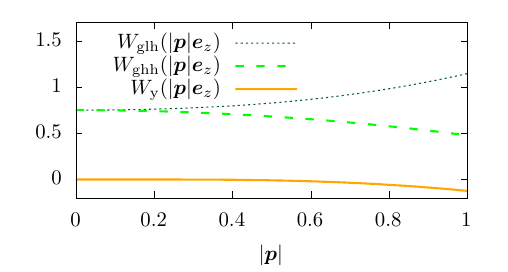}
  \caption{Energy surfaces for momentum aligned with the $z$ axis. The
    energy surfaces correspond to the green exciton with light holes
    (glh) and heavy holes (ghh) as well as the yellow series (y). The
    units are chosen such that $\hbar = e = m_{\mathrm{0}} / \gamma_1'
    = 1 / (4\pi\varepsilon_{\mathrm{0}} \varepsilon) = 1$.}
  \label{fig:surfaces_z}
\end{figure}

\begin{figure}
  \includegraphics[width=\columnwidth]{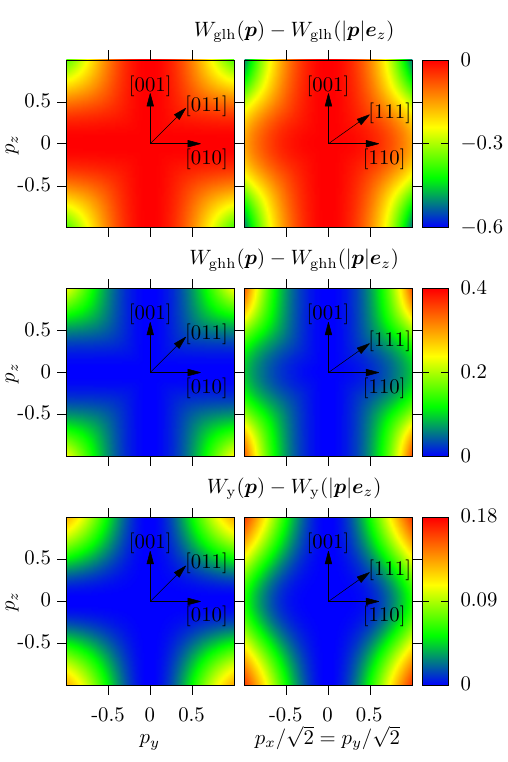}
  \caption{Visualization of the symmetry of the energy surfaces in the
    plane normal to $[100]$ (left) and normal to $[1\bar{1}0]$
    (right). The deviation from the corresponding energy surface with
    momentum aligned with the $z$ axis
    $W_k(\boldsymbol{p})-W_k(|\boldsymbol{p}|\boldsymbol{e}_z)$ is
    shown. The $C_{4v}$ symmetry of the plane normal to $[100]$ and
    the  $C_{2v}$ symmetry of the plane normal to $[1\bar{1}0]$
    becomes apparent.  The units are chosen such that $\hbar = e =
    m_{\mathrm{0}} / \gamma_1'   = 1 / (4\pi\varepsilon_{\mathrm{0}}
    \varepsilon) = 1$.}
  \label{fig:surfaces}
\end{figure}

Compared to the simple two-band model the
Hamiltonian~\eqref{eq:H_spins} now contains the additional spin
degrees of freedom 
and thus completely quantum mechanical quantities with no classical counterpart.
To arrive at a classical description we consider the timescale on which the dynamics
in the spin degrees of freedom and relative space take place, respectively.
Their corresponding timescale is characterized by the inverse of the
energy difference of adjacent levels.
While for the spin degrees of freedom the dominant contribution comes from the 
spin-orbit coupling term~\eqref{eq:H_SO} where the energy difference
stays constant at $\Delta=131$meV\cite{Schoene2016Deviations}, the
dominant contribution for the relative motion stems from the
hydrogen-like part for which the energy spacings shrink with $\sim 1/n^3$.
This situation is visualized in figure~\ref{fig:energy}.
Since the excitonic Rydberg energy $E_\mathrm{Ryd}=86$meV is in the same order of
magnitude as the spin-orbit splitting
the energy spacings become increasingly smaller for higher principle
quantum numbers, resulting in much slower dynamics for the relative
motion compared to the spin degrees of freedom.
In analogy to the Born-Oppenheimer approximation where the fast
electrons instantaneously react to a new position of the core we take
an adiabatic approach.
In case of cuprous oxide the fast motion of the spin degrees of
freedom reacts instantaneously to a new configuration in relative space.
For the spin degrees of freedom we diagonalize the Hamiltonian in 
a basis for quasi- and hole spin\cite{Ertl2020Classical}
\begin{equation}
  H_{\mathrm{band}} (\boldsymbol{p},\boldsymbol{I},\boldsymbol{S}_{\mathrm{h}}) \chi_k(\boldsymbol{p}; \boldsymbol{I},\boldsymbol{S}_\text{h} )= W_k(\boldsymbol{p}) \chi_k(\boldsymbol{p}; \boldsymbol{I},\boldsymbol{S}_\text{h} )\, .
\end{equation}
The corresponding eigenfunctions $\chi_k(\boldsymbol{p};
\boldsymbol{I},\boldsymbol{S}_\text{h} )$ and eigenvalues
$W_k(\boldsymbol{p})$ depend on the exciton momentum as a parameter.
This procedure leads to a description via three distinct energy
surfaces in momentum space $W_k(\boldsymbol{p})$, one for the yellow
series and two for the green series, for light and heavy holes respectively.
Similarly, the Luttinger parameters can be obtained by fitting the
model to full band structure calculations.\cite{Schoene2016Coupled,Schoene2016Deviations}
This means there is a one to one correspondence between the different
series and one of the energy surfaces respectively.
Figure~\ref{fig:surfaces_z} shows the three distinct energy surfaces
for momentum aligned with the $z$ axis.
The energy surfaces inherit the cubic symmetry of the system. 
This can be seen in figure~\ref{fig:surfaces} where the deviations
from the spherical symmetry are visualized in the two distinct
symmetry planes of the crystal.
With the adiabatic approach for the spin dynamics the cuprous oxide
Hamiltonian now takes the form
\begin{equation}
  H = E_{\mathrm{g}} + \frac{\gamma'_1}{2m_0} \boldsymbol{p}^2
  -\frac{e^2}{4\pi\varepsilon_0\varepsilon|\boldsymbol{r}|} + W_k(\boldsymbol{p}) \, .
\label{eq:H}
\end{equation}
The classical exciton dynamics can be investigated by choosing an
energy surface and integrating Hamilton's equations of motion of
the corresponding Hamilton function for a given energy.

\subsection{Methods for analyzing the phase space}
The classical equations of motion for excitons allow for either
fully three-dimensional orbits or two-dimensional orbits in one of the
symmetry planes (see figure~\ref{fig:symmetries}) of the crystal.
Here we briefly discuss methods for analyzing the corresponding six-
or four-dimensional phase space.

\subsubsection{\ac{PSOS}}
In case of a classical dynamics in a two-dimensional plane with
coordinates, e.g., $x$ and $y$ the corresponding four-dimensional
phase space can be analyzed by using a \ac{PSOS}.
Therefore, we choose the $y$ axis as 1D-sectional plane and
denote the coordinate $y$ and momentum $p_y$ whenever an orbit passes
the sectional plane.
By choosing $x=0$ the remaining momentum  $p_x$ is given by the
conservation of energy.
In the integrable case action angle variables can be constructed, and
the motion is confined to tori, leading to a completely regular phase
space.
When adding a small non-integrable perturbation, the KAM theorem states
that almost all non-resonant tori survive the perturbation.
When the perturbation grows more and more tori start to break up.
The dynamics becomes chaotic,
leading to stochastic areas in the
\ac{PSOS}.\cite{schuster2005deterministic,lichtenberg2013regular}
According to the Poincaré-Birkhoff theorem the resonant tori break up
leading to an equal number of elliptic and hyperbolic fixed points in
the \ac{PSOS} and eventually to the onset of chaos.\cite{schuster2005deterministic,lichtenberg2013regular}

\subsubsection{Stability analysis}
A more quantitative measure for the stability of individual trajectories
can be obtained by investigating Lyapunov exponents of orbits, which
describe the rate with which a deviation between neighboring
trajectories grows in the corresponding direction.
For periodic orbits with period $T$ it is given by
\begin{equation}
  \lambda_i =   \frac{1}{T} \ln (|d_i(0,T)|) \, ,
\label{eq:lambda}
\end{equation}
with $d_i(0,T)$ the eigenvalues of the stability matrix
$\boldsymbol{M} (0, T)$, which is a linearization for the time
evolution for deviations of the phase-space vector
$\boldsymbol{\gamma} = (\boldsymbol{r},\boldsymbol{p})^\top$, i.e.
\begin{equation}
  \Delta \boldsymbol{\gamma} (T) = \boldsymbol{M} (0, T) \Delta \boldsymbol{\gamma} (0) \, .
\end{equation}
In Hamiltonian systems the stability matrix $\boldsymbol{M}(0,T)$ of
periodic orbits is symplectic, and therefore the eigenvalues appear in
pairs $d_i$ and $1/d_i$.
For eigenvalues $d_i= e^{i \varphi}$ on the unit circle in the complex
plane the deviations of the phase-space vector of the corresponding
direction oscillate around the initial orbit.
The corresponding Lyapunov exponent is $\lambda_i=0$ and the direction is
called stable.
For any pair of eigenvalues $d_i$ not located on the unit circle an
unstable direction with positive Lyapunov exponent $\lambda_i>0$ does
exist.
The case $d_i=1$ indicates a marginally stable direction, which occurs
exactly at the bifurcation point of an orbit, or when a conserved
quantity, e.g., the energy in autonomous Hamiltonian systems, does
exist.
In the following discussion we omit the marginally stable directions
and only consider the stability properties in directions orthogonal to
the orbit.

To obtain the Lyapunov exponents we integrate the stability matrix along the
periodic orbit.
For  Hamiltonian systems its equation of motion is given by
\begin{equation}
  \frac{\mathrm{d}}{\mathrm{d}t} \boldsymbol{M}
  = \boldsymbol{J} \frac{\partial^2 H}{\partial \boldsymbol{\gamma^2}}\boldsymbol{M} \, ,
\end{equation}
with the symplectic matrix
\begin{equation}
\boldsymbol{J}=
\begin{pmatrix}
0 & I_n \\
-I_n & 0 \\
\end{pmatrix} 
~~\text{and}~~ \boldsymbol{M}(t=0)=
\begin{pmatrix}
 I_n & 0 \\
 0 & I_n  \\
\end{pmatrix}
\, ,
 \end{equation}
where $I_n$ is the $n\times n$ identity matrix.

\subsubsection{\acp{LD}}
While the dynamics in a two-dimensional plane can be
visualized through a \ac{PSOS}, this method is no longer well
suited for the analysis of fully three-dimensional or even
higher-dimensional motion.
One possible method to gain a better understanding of the phase space
is to investigate the dynamics using an
\ac{LD}.\cite{mendoza2010hidden,mancho2013lagrangian}
It is defined as the arc length of a trajectory with given initial
conditions $\boldsymbol{\gamma}_0$
\begin{equation}
  \mathrm{LD}\left(\boldsymbol{\gamma}_0,T\right)=\int_0^T ||\boldsymbol{v}|| \mathrm{d}t \, ,
\end{equation}
in a given time interval $[0,T]$.
The general idea is that in regular regions, where a small change of the
initial conditions leads to a qualitatively similar picture, the \ac{LD}
also will not change much, whereas one would expect stronger changes
when entering a phase space region with qualitatively different
behavior.
While this already provides a useful tool for obtaining an overview
over the general global phase space structures, the challenge of this
approach is that locally the detailed phase space structures cannot
be well resolved, when the arc length has a strong dependence on
the initial conditions of the orbit.
To gain a better resolution of the local fluctuations, recently
an indicator for chaos based on the second derivatives of
the \ac{LD} with respect to the initial conditions $\gamma_{0i}$ has
been introduced,\cite{Daquin2022Global} which reads
\begin{equation}
  \Delta \mathrm{LD}
  = \sum_{i=1}^N \left|\frac{\partial^2\mathrm{LD}}{\partial \gamma_{0i}^2}\right| \, .
\label{eq:DLD}
\end{equation}
Regions with large values of the $\Delta \mathrm{LD}$
indicator reveal chaotic dynamics, while low values with smooth
behavior indicate regular phase space regions.

\section{Results and discussion}
\label{sec:results}
To obtain classical exciton orbits we numerically integrate Hamilton's
equations of motion
\begin{equation}
  \dot r_i = \frac{\gamma'_1}{m_0} p_i + \frac{\partial
              W_k(\boldsymbol{p})}{\partial p_i} \; , \quad
  \dot p_i = -\frac{e^2}{4\pi\varepsilon_0\varepsilon}
              \frac{r_i}{|\boldsymbol{r}|^3} \, , 
\end{equation}
resulting from the Hamiltonian~\eqref{eq:H} by
using a Runge-Kutta method with adaptive step size.\cite{Brankin1993RKSUite}
To characterize the energy we use $n_\mathrm{eff}=
\sqrt{E_{\mathrm{Ryd}}/(E_{\mathrm{g}}-E)}$ for the yellow series and
$n_\mathrm{eff}= \sqrt{E_{\mathrm{Ryd}}/(E_{\mathrm{g}}+\Delta-E)}$
for the green series, which in the hydrogen-like case would correspond
to the principle quantum number of yellow and green series
respectively. For the calculations we use the same material parameters
as in reference~\onlinecite{Ertl2024}.

\subsection{\ac{PSOS}}
We now investigate the classical exciton dynamics for a
two-dimensional motion of the exciton in one of the two different
symmetry planes of the crystal shown in figure~\ref{fig:symmetries}.
The coordinates in the plane are then $y$ and $z$ for the plane
perpendicular to the $[100]$ axis, and $\sqrt{2}x=\sqrt{2}y$ and $z$
for the plane perpendicular to the $[1\bar 10]$ axis, respectively.
For convenience, we use a scaling of the coordinates and momenta,
$\boldsymbol{r} = n_{\mathrm{eff}}^2\tilde{\boldsymbol{r}}$,
$\boldsymbol{p} = n_{\mathrm{eff}}^{-1}\tilde{\boldsymbol{p}}$
in the figures.

\subsubsection{Dynamics at $n_{\mathrm{eff}}=5$}
\begin{figure*}
 \includegraphics[width=\textwidth]{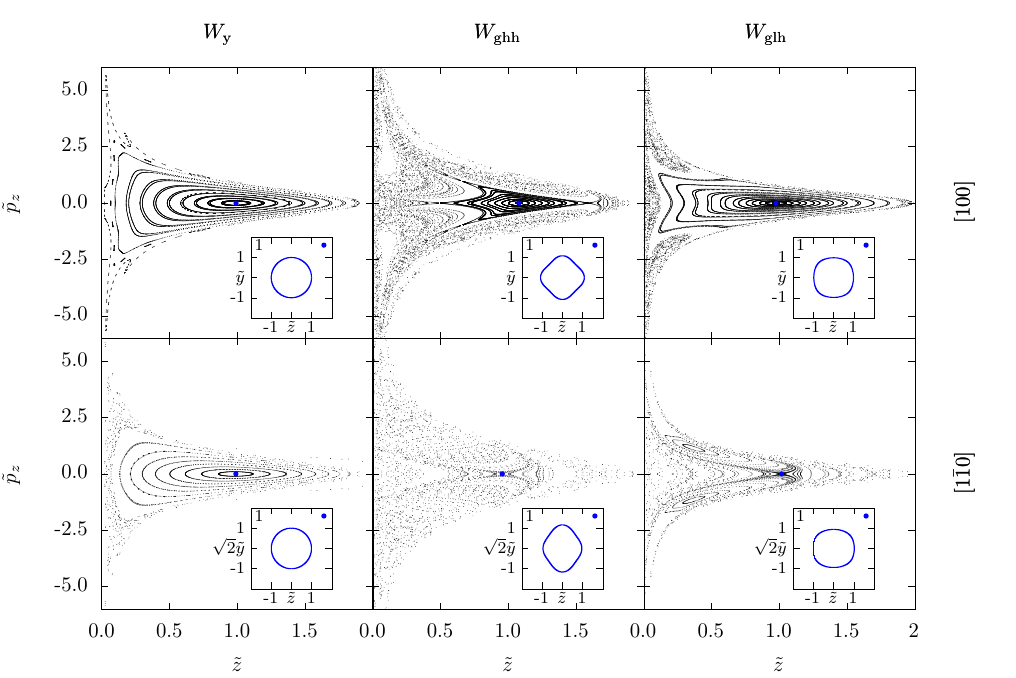}
  \caption{\ac{PSOS} at $n_\mathrm{eff}=5$ for the yellow exciton
    series (left), the green exciton series with heavy holes (middle),
    and the green exciton series with light holes (right) for the
    plane normal to $[100]$ (top), and the plane normal to
    $[1\bar{1}0]$ (bottom).  Insets show the orbit for the central
    fixed point. The units are chosen such that $\hbar = e =
    m_{\mathrm{0}} / \gamma_1' = 1 / (4\pi\varepsilon_{\mathrm{0}} \varepsilon) = 1$.
    Coordinates and momenta are scaled according to $\boldsymbol{r} =
    n_{\mathrm{eff}}^2\tilde{\boldsymbol{r}}$, 
    $\boldsymbol{p} = n_{\mathrm{eff}}^{-1}\tilde{\boldsymbol{p}}$.}
  \label{fig:comparison}
\end{figure*}

In figure~\ref{fig:comparison} \acp{PSOS} for $n_{\mathrm{eff}}=5$ for
both symmetry planes are presented for the yellow series and the green
series with heavy and light hole.
For the dynamics in the plane normal to $[100]$ one can observe a
central elliptical fixed point surrounded by regular tori.
For the yellow exciton series the majority of phase space exhibits
regular structures, with only a small chaotic region in the outermost
part of the \ac{PSOS}, while for the green series the chaotic regions
are larger and the tori in the outermost part of the \ac{PSOS} appear
stronger deformed then in the case for the yellow series.
The orbit for the central fixed point appears
almost circular for the yellow series in the plane normal to $[100]$
as well as in the plane normal to $[1\bar{1}0]$ resembling the
circular orbit one would obtain for similar starting conditions in the
hydrogen-like model. By contrast, the central fixed points for the
green series show stronger deviations from the hydrogen-like behavior,
instead one can observe the $C_{4v}$ symmetry for the orbit in the
plane normal to $[100]$ and the  $C_{2v}$ symmetry for the orbit in
the plane normal to $[1\bar{1}0]$ respectively. Concerning the phase
space in the symmetry plane normal to $[1\bar{1}0]$ the emerging
structures for the yellow series look fairly similar to what can be
observed in the plane normal to $[100]$ with a central elliptical
fixed point surrounded by a large region with regular tori and a small
chaotic region in the outermost part of the \ac{PSOS}.
For the green series the structures, however, differ entirely.
Here, the central periodic orbit becomes a hyperbolic fixed point
surrounded by a stochastic region in the \ac{PSOS} indicating
  deterministic chaos in phase space.
For the green series with heavy
holes the remaining part of the phase appears entirely chaotic while
for the green series with light holes some regular structures persists
surrounding two elliptical fixed points in addition to the central
hyperbolic fixed point.

\subsubsection{Bifurcations of  the central green exciton orbit}
\begin{figure}
  \includegraphics[width=0.9\columnwidth]{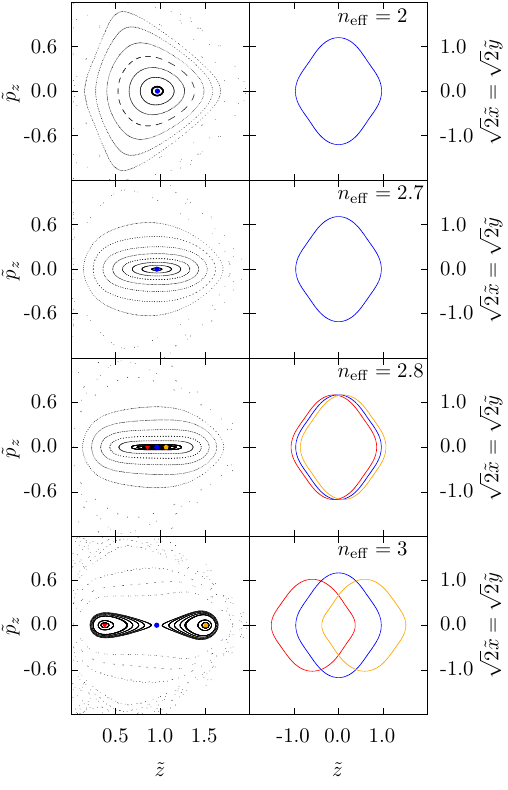}
  \caption{\ac{PSOS} for the plane normal to $[1\bar{1}0]$ for the green
    series with heavy holes (left) and orbits corresponding to the 
    observable fixed points (right). The energy ranges from values 
    corresponding to effective principle quantum numbers 
    $n_\mathrm{eff}=2$, $n_\mathrm{eff}=2.7$, $n_\mathrm{eff}=2.8$ to 
    $n_\mathrm{eff}=3$ from top to bottom. The units are chosen such
    that $\hbar = e = m_{\mathrm{0}} / \gamma_1'
    = 1 / (4\pi\varepsilon_{\mathrm{0}} \varepsilon) = 1$.
    Coordinates and momenta are scaled according to $\boldsymbol{r} =
    n_{\mathrm{eff}}^2\tilde{\boldsymbol{r}}$,
    $\boldsymbol{p} = n_{\mathrm{eff}}^{-1}\tilde{\boldsymbol{p}}$.}
  \label{fig:bifurcation}
\end{figure}

For the yellow exciton series the phase space in the two distinct symmetry
planes behave similar when varying the energy.\cite{Ertl2024} 
For the green series this holds true for the plane normal to $[100]$,
a closer investigation of the energy dependence of the green series 
reveals however, that the central orbit in the plane normal to $[1\bar{1}0]$
undergoes a bifurcation going from an elliptical fixed point,
splitting into a hyperbolic fixed point and two elliptical orbits.
We present this process for the green series with heavy holes,
the investigation of the light hole yields similar results.
The \acp{PSOS} and the shape of the central orbit at various values of
$n_{\mathrm{eff}}$ in the range from $n_{\mathrm{eff}}=2$ to $3$ are presented
in figure~\ref{fig:bifurcation}.
For energies corresponding to principle quantum number up to  $n_\mathrm{eff}=2.7$ 
the \ac{PSOS} reveals a stable elliptical fixed point, at $n_\mathrm{eff}=2.8$ 
the structure has changed now showing an unstable fixed point at the center 
of the \ac{PSOS} as well as two separated stable elliptical fixed points 
which move outward with increasing energy. The orbits corresponding to 
the fixed points are shown below their respective \ac{PSOS}. Up to 
$n_\mathrm{eff}=2.7$ there is only one orbit with the full 
$C_{2\mathrm{v}}$ symmetry of the $[1\bar{1}0]$ plane. This still holds 
for the unstable orbit of the central fixed point, however the two 
orbits corresponding to the elliptical fixed point no longer exhibit 
the full symmetry of the plane. Instead, they can be transformed into 
each other using symmetry operations of the $C_{2\mathrm{v}}$ group.
In the full coordinate space
the central fixed point undergoes a pitchfork bifurcation, changing its stability in the plane from stable to unstable.
This can also be viewed as a period doubling in the fundamental domain:
The cubic $O_{\mathrm{h}}$ symmetry of cuprous oxide implies that the dynamics
can be characterized in the symmetry-reduced fundamental domain.
The dynamics in the full coordinate space is then obtained by applying
symmetry operations from $O_{\mathrm{h}}$.
For the plane normal to $[1\bar{1}0]$ there are two mirror axes which
correspond to coordinate axes in figure~\ref{fig:bifurcation} (right).
Using these axes to project the orbits into the fundamental domain
(top right quadrant) the orbits corresponding to the stable fixed
points coincide, which means that the period of the central fixed
point in the fundamental domain is half the period of the orbit in the
full coordinate space.
By contrast, the orbit corresponding to the stable fixed points has the
same period in the fundamental domain and in the full coordinate space.
As a consequence, in the fundamental domain the initially stable orbit
splits into an unstable orbit with the same period and a stable orbit
with twice the period, i.e., the orbit undergoes a period-doubling
bifurcation with increasing energy.

\begin{figure}
  \includegraphics[width=0.9\columnwidth]{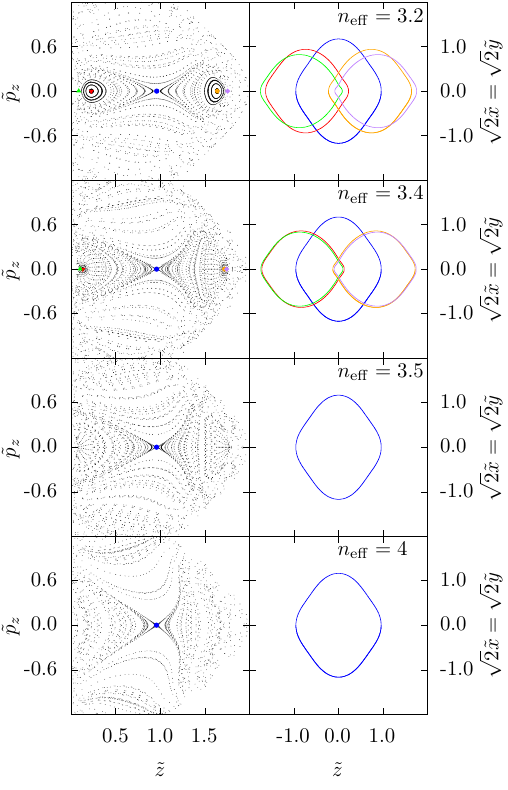}
  \caption{\ac{PSOS} for the plane normal to $[1\bar{1}0]$ for the green
    series with heavy holes (left) and orbits corresponding to the 
    observable fixed points (right). The energy ranges from values 
    corresponding to effective principle quantum numbers 
    $n_\mathrm{eff}=3.2$, $n_\mathrm{eff}=3.4$, $n_\mathrm{eff}=3.5$ to 
    $n_\mathrm{eff}=4$ from top to bottom. The units are chosen such
    that $\hbar = e = m_{\mathrm{0}} / \gamma_1'
    = 1 / (4\pi\varepsilon_{\mathrm{0}} \varepsilon) = 1$.
    Coordinates and momenta are scaled according to $\boldsymbol{r} =
    n_{\mathrm{eff}}^2\tilde{\boldsymbol{r}}$,
    $\boldsymbol{p} = n_{\mathrm{eff}}^{-1}\tilde{\boldsymbol{p}}$.}
  \label{fig:bifurcation_2}
\end{figure}

For the green series with light holes a similar behavior can be
observed. In contrast to the series with heavy holes the split off
stable fixed points do not stay on the $p_z=0$ axis instead they move
along the $p_x=p_y=0$ axis.
The corresponding fixed points can still be seen in figure~\ref{fig:comparison}.
For higher values the elliptical fixed points move outward towards the
chaotic region showing a clear separation for example for
$n_\mathrm{eff}=3$ in figure~\ref{fig:bifurcation} for the green
series with heavy holes.
Increasing the energy further the elliptical fixed points formed in the
pitchfork bifurcation move further towards the chaotic region.
There they finally collide with a hyperbolic fixed point and disappear
in a saddle-node bifurcation. This scenario is depicted in 
figure~\ref{fig:bifurcation_2}.
This results in a completely chaotic phase
space, as can be seen for the green series with heavy holes at
$n_\mathrm{eff}=5$ in figure~\ref{fig:comparison}.

\subsection{Lyapunov exponents}
\begin{figure}
\includegraphics[width=0.92\columnwidth]{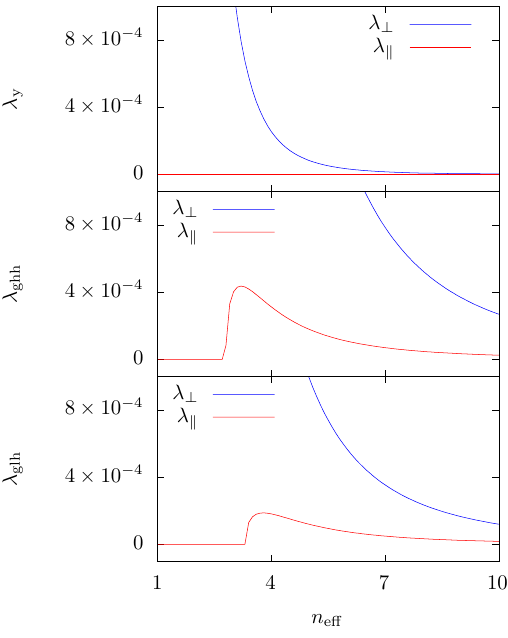}
  \caption{Lyapunov exponents for the orbits corresponding to the
    central fixed points in the plane normal to $[1\bar{1}0]$ for the
    yellow series (top), the green series with heavy holes (middle),
    and the green series with light holes (bottom). The energy ranges
    from values corresponding to effective principle quantum numbers
    $n_\mathrm{eff}=1$ to $n_\mathrm{eff}=10$. The units are chosen
    such that $\hbar = e = m_{\mathrm{0}} / \gamma_1' = 1 /
    (4\pi\varepsilon_{\mathrm{0}} \varepsilon) = 1$.}
  \label{fig:Lyapunov}
\end{figure}

The stability of the central fixed points can also be analyzed
quantitatively using Lyapunov exponents. In addition to the dynamics in 
the plane this also enables us to gain insights concerning the 
stability of the orbit against perturbations out of the plane.
For the plane normal to $[100]$ the central fixed points of all series 
are stable both against perturbations in the plane, resulting in the
elliptical  shape of the fixed point in figure~\ref{fig:comparison},
and for  perturbations out of the symmetry plane.
Thus, all Lyapunov exponents are equal to zero in this case.
For the plane normal to $[1\bar{1}0]$ the situation becomes 
more complex.
The Lyapunov exponents for the orbits corresponding to the central
fixed points in the plane normal to $[1\bar{1}0]$ are shown in
figure~\ref{fig:Lyapunov}.
Note that the small values of the Lyapunov exponents ($\lambda_i\sim 10^{-4}$)
are due to the fact that the logarithm of the eigenvalues in equation~\eqref{eq:lambda}
are to be divided by the period, which in a hydrogen-like system scales with the
energy according to $T= 2\pi n_\mathrm{eff}^3$.
It becomes apparent, that the Lyapunov exponent corresponding to
perturbations of the orbit out of the symmetry plane $\lambda_\perp$
shows unstable behavior for all energies, i.e.,
the Lyapunov exponents are positive and approach zero for larger energies.
In this case the dominant contribution from the band structure
corrections is due to the spin-orbit coupling, which then results in a
more hydrogen-like behavior of the three series.
In the hydrogen-like case the central orbit would be a circle with all
Lyapunov exponents equaling zero, since this case is completely regular.
For the yellow series the central orbit bares resemblance to a circle,
and also the Lyapunov exponent corresponding to perturbations in the
symmetry plane $\lambda_\parallel$ shows stable behavior for all given
energies.
By contrast, the central orbits corresponding to the green series show
stronger deviations from the circular shape. In addition, their
stability properties in the plane differ from the hydrogen-like
behavior. While the Lyapunov exponents are equal to zero for low
energies a change in stability can be observed for both green series. 
For the series with heavy holes this happens between 
$n_\mathrm{eff}=2.7$ and $n_\mathrm{eff}=2.8$, matching the transition
observed in figure~\ref{fig:bifurcation}.
For the series with light holes this change can be observed
between $n_\mathrm{eff}=3.3$ and $n_\mathrm{eff}=3.4$. After changing
their stability the Lyapunov exponents start decreasing again and
approach $\lambda_\perp=0$ at high energies.

\subsection{Analysis of the full phase space with \acp{LD}}
\begin{figure}
\includegraphics[width=0.95\columnwidth]{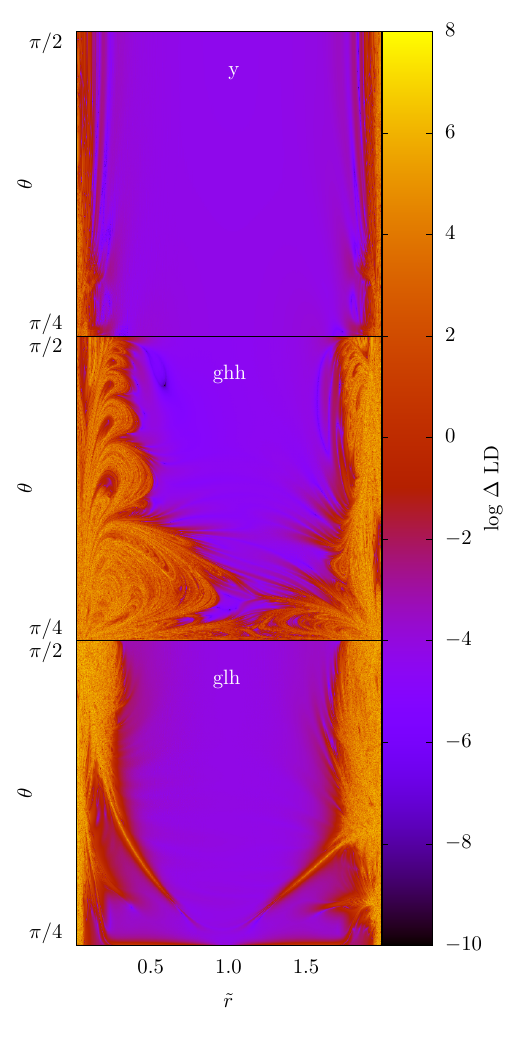}
\caption{$\Delta \mathrm{LD}$ at $n_{\mathrm{eff}}=5$ for the yellow
  exciton series (top), the green exciton series with heavy holes
  (middle), and the green exciton series with light holes (bottom).
  At the upper boarder the plane normal to $[001]$ and at the lower
  boarder the plane normal to $[\bar{1}01]$ are located. Strong
  fluctuations in the arc length are characterized by large values
  of $\Delta \mathrm{LD}$ and indicate a classically chaotic exciton
  dynamics. The chaotic regions are more pronounced for the green than
  for the yellow excitons.}
  \label{fig:LD}
\end{figure}
In cuprous oxide periodic three-dimensional orbits do exist when
moving the initial conditions out of the symmetry planes.
The two-dimensional periodic orbits exhibit a turning point in the
radial direction on one of the symmetry axis ($[100]$, $[1\bar{1}0]$
or equivalents).
The projection of the three-dimensional periodic orbits on the
symmetry plane exhibit the same behavior with either the velocity
exhibiting an angle to the plane or the velocity parallel of the plane
and the starting position shifted above the symmetry
axis.\cite{Ertl2024}
This allows for the characterization of the three-dimensional periodic
orbits by only the initial radius, i.e., the $r$ coordinate and one of
these angles.
Changing these parameters, however, in general results in non-periodic
three-dimensional orbits.
We use this property to examine the three-dimensional motion for the three 
distinct exciton series.
To examine the three-dimensional motion we initialize the orbit at
$\varphi=0$ in a turning point for the $r$ and $\vartheta$ direction
exemplarily.
We thus can visualize the three-dimensional dynamics using the
$\Delta\mathrm{LD}$ indicator~\eqref{eq:DLD} over $r$ and $\vartheta$
calculating the \ac{LD} on a $500\times500$ mesh and using a finite
difference quotient for the second derivatives as in
reference~\onlinecite{Daquin2022Global} and a time $T=20 \pi n_\mathrm{eff}^3$
corresponding to 10 cycles in a hydrogen-like model.
The results are presented in figure~\ref{fig:LD}.
At the values $\vartheta=\pi/2$ and $\vartheta=\pi/4$ at the 
top and bottom of figure~\ref{fig:LD} the initial 
conditions result in two-dimensional motion in the symmetry planes normal
to  $[001]$ and $[\bar{1}01]$ respectively. These symmetries planes are
equivalent to the symmetries planes normal to $[100]$ and $[1\bar{1}0]$
depicted in figure~\ref{fig:comparison}. 
In addition, the \ac{LD} indicator allows to investigate the global 
phase-space structure for three-dimensional motion in the transition 
region between the planes.

Regions with large values of $\Delta \mathrm{LD}$ indicate
chaotic behavior, whereas smaller values indicate regions with regular
behavior.
For the yellow exciton series chaotic regions only exist at the lowest
and highest $r$ values, with large regular regions in between.
For the two green exciton series the situation is similar close to the
plane normal to $[001]$ but the area of the chaotic regions increases
when the plane normal to $[\bar{1}01]$ is approached.
Here, the $\Delta\mathrm{LD}$ indicator exhibits a completely chaotic
phase space for the green series with heavy holes and a small regular
region remaining for the green series with light holes at $n_{\mathrm{eff}}=5$
matching the appearance of the \ac{PSOS} in figure~\ref{fig:comparison}.
Figures for the $\Delta \mathrm{LD}$ indicator, where the starting
conditions are chosen to match the other periodic three-dimensional
orbits, result in qualitatively similar pictures.
The same holds true when varying $n_{\mathrm{eff}}$, the emerging 
phase space structures are qualitatively similar to $n_{\mathrm{eff}}=5$.
While for the green series the share of the phase space which exhibits
chaotic regions grows slightly, for the yellow series, 
however, the chaotic regions shrink with increasing
$n_{\mathrm{eff}}$ resulting in a more regular phase space.

\section{Conclusion and outlook}
\label{sec:conclusion}
We have analyzed the classical dynamics of excitons in cuprous oxide 
using an adiabatic approach considering the complex band structure of 
the crystal. The description via energy surfaces in momentum space 
allows to individually investigate the dynamics of the corresponding 
exciton series. Using \acp{PSOS} and \acp{LD} we demonstrated that the 
classical dynamics of the yellow exciton series is mostly regular,
while for the green series with light and heavy holes large chaotic
regions exist. Excitons in cuprous oxide thus provide an example of 
a two-particle system with classical chaos, even without the introduction
of external fields. The chaotic dynamics arises due to the complex 
valence band structure of the system.

In principle the classically chaotic dynamics should manifest in the 
level spacing statistics.
While the level spacing statistics of integrable systems is given by a
Poissonian distribution the level spacing statistics of classically
chaotic systems can be described by random matrix
theories.\cite{Haake2018Quantum}
Because of the large chaotic regions in the classical dynamics of the
green exciton series we expect significant deviations in the
corresponding level-spacing distribution from the Poissonian
distribution of a regular system.

It would also be interesting to study the impact of the observed bifurcation 
on the quantum spectra by analyzing the quantum recurrence spectra of
the green series similar to the yellow exciton series,\cite{Ertl2022Signatures,Ertl2024}
where peaks in the quantum recurrence spectra occur at the action of a
corresponding orbit.
After the bifurcation takes place one should be able to identify a splitting 
of the corresponding peak in the quantum recurrence spectrum. 

Due to the coupling to the yellow continuum, the green exciton states
are resonances with finite lifetimes.\cite{Rommel2020Green}
The decay of these states is related to a surface hopping between the
green and yellow energy surfaces.
This surface hopping may be modeled within the adiabatic approach when
additional non-adiabatic coupling terms are considered, similar as for
the surface hopping on molecules within the Born-Oppenheimer
approximation.\cite{Tully1990Molecular} 
In cuprous oxide this would allow the exciton to jump between green 
and yellow energy surfaces, until it reaches the unbound region on the
yellow energy surface where it dissipates.
This extension of the adiabatic approach would allow for a
semiclassical interpretation of the resonant states in cuprous oxide.
The adiabatic approach will also allow for studying the classical dynamics
of magnetoexcitons, viz.\ a system, where all antiunitary symmetries
are broken.\cite{Schweiner2017MagnetoexcitonsBreak,Schweiner2017goe,schweiner2017crossover}

\begin{acknowledgments}
This work was supported by Deutsche Forschungsgemeinschaft (DFG)
through SPP 1929 GiRyd and Grant No.~MA1639/16-1.
\end{acknowledgments}

\section*{AUTHOR DECLARATIONS}
\subsection*{Conflict of Interest}
The authors have no conflicts to disclose.
\subsection*{Author Contributions}
\textbf{Jan Ertl:} Formal analysis, Investigation, Methodology,
Software, Supervision, Validation, Visualization, Writing – original
draft, Writing – review \& editing.
\textbf{Sebastian Rentschler:} Formal analysis, Investigation,
Software, Writing – original draft, Writing – review \& editing.
\textbf{Jörg Main:} Conceptualization, Funding acquisition,
Methodology, Project administration, Resources, Supervision, Writing –
original draft, Writing – review \& editing.

\subsection*{DATA AVAILABILITY}
The data that support the findings of this study are available
from the corresponding author upon reasonable request.

\section*{References}
%

\end{document}